\documentclass[12pt]{article}
\usepackage[dvips]{graphicx}

\parskip=\medskipamount

\def\be{\begin{equation}}
\def\ee{\end{equation}}
\def\bea{\begin{eqnarray}}
\def\eea{\end{eqnarray}}
\def\bean{\begin{eqnarray*}}
\def\eean{\end{eqnarray*}}
\def\r#1{(\ref{#1})} 
\def\la#1{\label{#1}}                                     
\def\c#1{\cite{#1}}
\def\b{\beta}
\def\D{\Delta}
\def\l{\lambda}
\def\t{\theta}

\title{Loop Groups and Discrete KdV Equations}
\author{Jeremy Schiff \\
        Department of Mathematics,  Bar-Ilan University\\
        Ramat Gan 52900, Israel\\
        email: schiff@math.biu.ac.il}
\date{September 2002}

\begin{document}

\maketitle

\begin{abstract}
A study is presented of fully discretized lattice equations associated 
with the KdV hierarchy.
Loop group methods give a systematic way of constructing discretizations 
of the equations in the hierarchy. The lattice KdV system of 
Nijhoff {\em et al.} arises from the lowest order discretization of the 
trivial, lowest order equation in the hierarchy, $b_t=b_x$. Two new
discretizations are also given, the lowest order discretization of the first nontrivial 
equation in the hierarchy, and a ``second order'' discretization of 
$b_t=b_x$. The former, which is given the name {\em full lattice KdV}
has the (potential) KdV equation as a standard continuum limit. For 
each discretization a B\"acklund transformation is given and soliton content 
analyzed. The full lattice KdV system has, like KdV itself,
solitons of all speeds, whereas  both other discretizations
studied have a limited range of speeds (being discretizations of 
an equation with solutions only of a fixed speed). 
\end{abstract}

\section{Introduction}

Despite the fact that numerical simulations of PDEs of KdV type can 
be done quickly and accurately these days using standard spectral methods, 
it is still of interest to look at discretizations of such PDEs, and 
see how ``integrability properties'' (elastic soliton scattering, 
existence of conserved quantities etc.) are  affected by 
discretization, and in particular to see if there are
``integrable discretizations'', that exhibit all the special 
properties of the underlying PDE. One can consider both ``partial''
and ``full'' discretizations; in the former only the spatial 
coordinate is discretized, in the latter time is discretized too.
This paper focuses on full discretizations.

The difference equation usually known as {\em discrete KdV}
was first studied by Hirota \c{Hirota}. Using a slightly different 
notation from that of \c{Hirota}, discrete KdV is the equation
\be
\frac1{1+u_{n+1,m+1}} - \frac1{1+u_{n,m}} = 
c(u_{n+1,m} - u_{n,m+1}) \qquad (c~{\rm constant})\ .
\la{dh}
\ee
This is a discretization of KdV, but in a rather unusual sense. The main 
justification for the name ``discrete KdV'' is that \r{dh} has a bilinear
formulation and a family of soliton solutions very similar to those 
of KdV (see also \c{hc} for rational solutions).  The study of 
discrete KdV was taken further by Nijhoff and collaborators 
(see \c{Nijhoff} for a review and references). 
The work of Nijhoff {\em et al.} focuses on the equation 
\be
\left( 1 + \frac{b_{n,m+1}-b_{n+1,m}}{p-q} \right)
\left( 1 + \frac{b_{n,m}-b_{n+1,m+1}}{p+q} \right) = 1 
\qquad (p,q~{\rm constant})\ ,
\la{dp}
\ee
which they call {\em lattice KdV}. In fact this equation is a 
``potential form'' of discrete KdV, in the sense
that if $b_{n,m}$ satisfies \r{dp}, then it is easy to check that 
\be u_{n,m} = \frac{b_{n-1,m-1}-b_{n,m}}{p+q}  \ee
satisfies \r{dh} with $c=(p+q)/(p-q)$. (This is supposed
to be an analog of
the fact that if $b(x,t)$ satisfies the ``potential KdV''
equation $b_t = \frac14 b_{xxx} + \frac32 b_x^2 +\delta(t) $
for some function $\delta(t)$, then 
$u=b_x$ satisfies the KdV equation $u_t=\frac14 u_{xxx} + 3uu_x$.)

Nijhoff {\em et al.}'s lattice KdV equation has an advantage over
Hirota's discrete KdV in that it is easier to see its continuum limit
(in the usual sense, to be explained shortly)
as well as at least one
nonstandard continuum limit in which it reduces
to the potential KdV equation. On substituting  $p=1/h$ and $q=1/k$,
\r{dp} becomes
\bea
-\frac{ b_{n+1,m+1}-b_{n,m+1}+b_{n+1,m}-b_{n,m} }{h}
&+&\frac{ b_{n+1,m+1}-b_{n+1,m}+b_{n,m+1}-b_{n,m} }{k} \nonumber\\
&+& (b_{n,m}-b_{n+1,m+1})(b_{n,m+1}-b_{n+1,m}) = 0\ . 
\la{dp2}
\eea
Taking the standard continuum limit will be taken to mean replacing 
$b_{n,m}$ by $b(x,t)$, $b_{n+1,m}$ by $b(x+h,t)$, $b_{n,m+1}$ by $b(x,t+k)$, 
$b_{n+1,m+1}$ by $b(x+h,t+k)$, expanding in powers of $h$ and $k$ and 
ignoring all but leading order terms. It is clear that in this limit
the first term in \r{dp2} gives $-2b_x$, the second $2b_t$ and the third $0$.
Thus {\em in the standard continuum limit, 
lattice KdV is simply a discretization of $b_t=b_x$}.   
A nonstandard continuum limit of \r{dp2} that gives 
the potential KdV equation is as follows: Make the same replacements as 
before, expand in powers of $h$ and $k$, but keep not only the leading order
terms but also all terms of order $h$ and $h^2$. This gives
\be
-2\left(b_x+\frac{h}{2}b_{xx}+\frac{h^2}{6}b_{xxx}\right) 
+b_t+\left(b_t+hb_{tx}+\frac{h^2}{2}b_{txx}\right) +h^2b_x^2 = 0 
\ee
Now write $b=\tilde{b}-\frac{h}{2}{\tilde{b}}_x$. 
Ignoring terms of order $h^3$ and above, the last equation can be written
\be
{\tilde{b}}_t={\tilde{b}}_x-\frac{h^2}{3}
\left(\frac14{\tilde{b}}_{xxx}+\frac32{\tilde{b}}_x^2\right)    \ .
\ee
This is a ``linear combination'' of the flow obtained in the 
standard continuum limit with the  potential KdV flow. 

The foregoing discussions raise a variety of questions. 
The relationship of KdV/potential KdV and discrete KdV/lattice KdV as it stands 
is rather cryptic and requires some clarification.
It would also be good to have another integrable lattice equation from which 
KdV/potential KdV can be obtained by taking a standard continuum 
limit. If this is possible, then it would be good to 
know just what freedoms there are in constructing integrable discretizations.
Finally, though this is a question that will not be addressed in the current
paper, given an integrable lattice equation, just how much freedom is
there in taking the continuum limit? 

This paper discusses the subject of discretizations of KdV 
using loop group methods. The basic fact behind the loop group 
approach to KdV is that the KdV equation (or, more precisely, the Lax
pair for the KdV equation) is simply a ``disguised'' version  of the 
Frobenius-integrable pair of linear first-order constant-coefficient ODEs
\be
\partial_x U = \pmatrix{ 0 & 1 \cr \l & 0 \cr} U \ , \qquad
\partial_t U = \l \pmatrix{ 0 & 1 \cr \l & 0 \cr} U 
\la{lin1} \ee 
(here $U$ is a $2\times 2$ matrix function of $x,t,\l$). The relation
of the above system with KdV will be explained fully in section 2
below. In greater generality, the $N$th flow 
($N=1,3,5,\ldots$) in the KdV hierarchy is associated with the system 
\be
\partial_x U = \pmatrix{ 0 & 1 \cr \l & 0 \cr} U \ , \qquad
\partial_{t_n} U =  \pmatrix{ 0 & 1 \cr \l & 0 \cr}^N U ,
\la{lin2}\ee
which reduces to the standard system \r{lin1} when $N=3$.  The 
approach proposed in this paper for constructing integrable discretizations
of KdV is simply to discretize the system \r{lin1} or \r{lin2} (any
explicit scheme for numerical integration of ODEs can be used) and then
to apply the necessary ``disguise'' to translate this system into 
a discrete KdV. Section 3 is devoted to the simplest
discretization of \r{lin2} with $N=1$, namely 
\be 
U_{n+1,m} = \left[ I + h\pmatrix{ 0 & 1 \cr \l & 0 \cr} 
\right] U_{n,m} , \qquad
U_{n,m+1} = \left[ I + k\pmatrix{ 0 & 1 \cr \l & 0 \cr} 
\right] U_{n,m} .
\la{disc1}
\ee
This is just a first-order Euler scheme with different step sizes in the 
$x$ and $t$ directions. This scheme gives rise to the lattice
KdV equation, which, as 
shown above, is a first-order discretization of the $N=1$ flow in
the potential KdV hierarchy, $b_t=b_x$. As an application of the loop group
formulation, a B\"acklund transformation for \r{dp} is given, 
and soliton solutions are derived (c.f. \c{Hirota}). 
A brief analysis of the soliton solutions is given, which 
helps clarify the rather schizophrenic nature of the lattice KdV 
equation, which on one hand is  a (nonlinear) discretization of 
$b_t=b_x$, and on the other displays features of potential KdV.  

Section 4 is devoted to the simplest discretization of \r{lin1}, namely
\be 
U_{n+1,m} = \left[ I + h\pmatrix{ 0 & 1 \cr \l & 0 \cr} 
\right] U_{n,m} , \qquad
U_{n,m+1} = \left[ I + k\l\pmatrix{ 0 & 1 \cr \l & 0 \cr} 
\right] U_{n,m} .
\la{disc2}
\ee
As expected, this gives rise to a system which is, in a natural way, 
a first-order discretization of the potential KdV equation. The
system is a little complicated, involving two auxiliary fields
(reminiscent of the discretization of the sinh-Gordon equation given in 
\c{bobenko}),  but it seems this is the price that has to be paid to have 
an integrable lattice equation that has potential KdV as a natural
continuum limit. The B\"acklund transformation and soliton 
solutions are derived for this system too. 

Section 5 considers another discretization of \r{lin2} for $N=1$, 
namely 
\bea 
U_{n+1,m} &=& \left[ I + h\pmatrix{ 0 & 1 \cr \l & 0 \cr} 
+ \frac{h^2}2 \pmatrix{ 0 & 1 \cr \l & 0 \cr}^2 
\right] U_{n,m} 
= \pmatrix{ 1+\frac{h^2\l}{2} & h \cr
            h \l & 1+\frac{h^2\l}{2} \cr} U_{n,m}\ , 
\la{so1}\\
U_{n,m+1} &=& \left[ I + k\pmatrix{ 0 & 1 \cr \l & 0 \cr} 
\right] U_{n,m} \ .
\la{so2}
\eea
This example is worked out mainly to illustrate that the method can
be extended to arbitrary order discretizations of \r{lin1} and \r{lin2}, 
establishing that there is quite
a lot of freedom in constructing integrable 
discretizations.
Section 6 contains some concluding remarks. 

I conclude the introduction with a brief mention of some relevant literature.  
The approach to discretization taken in this paper is closely related to 
the approach of discretizing the scattering problem, which was first 
proposed by  Ablowitz and Ladik \c{al}, and recently has been revisited by 
Boiti {\em et al.} \c{BPPS}. Several potentially interesting applications of 
discretizations of equations of KdV type have emerged recently. 
Nijhoff {\em et al.} \c{Nijhoff} were the first to notice the link 
between lattice KdV and the {\em discrete conformal map} equation 
\be
\frac{(z_{n,m}-z_{n+1,m})(z_{n,m+1}-z_{n+1,m+1})}
{(z_{n,m}-z_{n,m+1})(z_{n+1,m}-z_{n+1,m+1})}=s
\qquad (s~{\rm constant})\ ,
\la{dcm}
\ee
which in the case $s=-1$  is a natural discretization of the Cauchy-Riemann 
conditions. Techniques related to those of this paper have been applied 
to \r{dcm} in \c{dcmpaper}. \r{dcm} may well play a 
significant role in the field of numerical conformal mapping. 
Discretizations of KdV and related equations have also been shown 
to have a role in a variety of other numerical algorithms \c{numapps}.

\section{KdV as a linear constant-coefficient flow}

This section contains a summary of results from \c{me1}, relating
the (potential) KdV equation 
with the linear constant coefficient flow \r{lin1}. 
A rather more mathematical description can be found in \c{sw}.

The general solution
of \r{lin1} is 
\be
U(x,t,\l) = \exp\left( 
x \pmatrix{ 0 & 1 \cr \l & 0 \cr} +  t \l \pmatrix{ 0 & 1 \cr \l & 0 \cr} 
\right) U(0,0,\l)\ .
\ee
Assume that the function $U(0,0,\l)$ is defined for $|\l|=1$, and has 
nonzero determinant; in other words it is an element of the {\em loop group}
$LGL_2({\bf C})$ \c{PS}. Then evidently so is $U(x,t,\l)$. Now a typical element
$g(\l)$ of the loop group $LGL_2({\bf C})$ can be written as a product 
$S^{-1}(\l)Y(\l)$ where $Y(\l)$ is holomorphic for $|\l|<1$ and 
$S(\l)$ is holomorphic for $|\l|>1$ with $S(\infty)=I$. This is the so-called
{\em Birkhoff factorization theorem}, see \c{PS}, chapter 8. So let us 
write 
\be 
U(x,t,\l) = S^{-1}(x,t,\l) Y(x,t,\l)\ ,  
\la{fact}\ee
(with $Y$ holomorphic in $|\l|<1$, $S$ holomorphic in $|\l|>1$ 
and $S(x,t,\infty)=I$)
and let us try to find differential equations satisfied by the two 
``components'' $Y$ and $S$ of $U$. Substituting \r{fact} into \r{lin1}, 
mulitplying on the left by $S$ and on the right by $Y^{-1}$ gives
\be
-S_xS^{-1} + Y_xY^{-1} = S\pmatrix{ 0 & 1 \cr \l & 0 \cr} S^{-1}\ , \qquad
-S_tS^{-1} + Y_tY^{-1} = \l S\pmatrix{ 0 & 1 \cr \l & 0 \cr} S^{-1}\ .
\la{16}
\ee
Now, if
\be
S=I + \frac1{\l} \pmatrix{ a_1(x,t) & b_1(x,t) \cr
                           c_1(x,t) & d_1(x,t) \cr }
    + \frac1{\l^2} \pmatrix{ a_2(x,t) & b_2(x,t) \cr
                           c_2(x,t) & d_2(x,t) \cr }
    + \ldots
\la{sexp}\ee
then a brief calculation shows 
\bea 
S\pmatrix{ 0 & 1 \cr \l & 0 \cr} S^{-1}\ 
&=& 
\pmatrix{ -b & 1 \cr
          \l-v & b \cr}
+ O\left(\l^{-1}\right)\ ,
\la{18}\\
\l S\pmatrix{ 0 & 1 \cr \l & 0 \cr} S^{-1}\ 
&=& 
\pmatrix{ -b \l - B  & \l +v-b^2 \cr
          \l^2-v\l - V  & b \l + B  \cr}
+ O\left(\l^{-1}\right)\ ,
\la{19}\eea
where  $b=-b_1$, $v=a_1-d_1$, 
$B=c_1-b_2+a_1b_1$, 
$V=a_2-d_2+a_1d_1-b_1c_1-a_1^2$. 
Substitute these results in \r{16}. 
On the left-hand side of the equations in \r{16}, since $Y$ is holomorphic
in $|\l|<1$, $Y_xY^{-1}$ and  $Y_tY^{-1}$ can be written as power series in 
$\l$. And since $S$ is holomorphic
in $|\l|>1$ with $S(x,t,\infty)=I$, $S_xS^{-1}$ and  $S_tS^{-1}$ can be written 
as power series in $1/\l$ with no constant term. Thus from the non-negative 
powers of $\l$ in \r{16}, after substituting \r{18}-\r{19}, it follows that 
\bea 
Y_x Y^{-1} &=& \pmatrix{ -b & 1 \cr
          \l-v & b \cr}\ ,   \la{lp1}   \\
Y_t Y^{-1} &=& \pmatrix{ -b \l - B  & \l +v-b^2 \cr
          \l^2-v\l - V  & b \l + B  \cr} \ . \la{lp2}
\eea 
If $X=Y_x Y^{-1}$, $T=Y_t Y^{-1}$ then $X,T$ must satisfy 
the zero-curvature equation
\be 
X_t - T_x + [X,T] = 0\ . 
\ee
Substituting the forms \r{lp1}-\r{lp2} into the  zero-curvature equation,
required to be true for all $\l$, gives the following system of equations:
\bea
v &=& b_x+b^2\ , \\
B &=& {\textstyle{\frac12}}b_{xx} +bb_x\ , \\ 
V_x &=& \left({\textstyle{\frac14}}b_{xxx}+{\textstyle{\frac12}} b_x^2 
        +bb_{xx}+b^2b_x\right)_x \ , \\
b_t &=& {\textstyle{\frac12}} b_{xxx} +2b_x^2+bb_{xx} +b^2b_x - V\ .
\eea
The third equation can be integrated to give
$V = \frac14b_{xxx}+bb_{xx}+\frac12 b_x^2 +b^2b_x -  \delta(t)$, 
where $\delta$ is an arbitrary function of $t$ alone. Using 
this in the last equation gives
\be 
b_t = {\textstyle{\frac14}} b_{xxx} + {\textstyle{\frac32}}b_x^2  + \delta(t) \ . 
\la{pkdv}\ee
All this can be summarized in the following result:

\smallskip

\noindent{\bf Proposition 2.1:} 
Let $U(0,0,\l)$ be an element of $LGL_2({\bf C})$, 
let $S^{-1}Y$ be the Birkhoff decomposition of 
$U(x,t,\l) = \exp\left( x \pmatrix{ 0 & 1 \cr \l & 0 \cr} 
+  t \l \pmatrix{ 0 & 1 \cr \l & 0 \cr} \right) U(0,0,\l)$, and let 
$b(x,t)$ be $(-1)$ times the component of $1/\l$ in the 1,2-entry of $S$. Then $b(x,t)$
is a solution, possibly with singularities, of
the potential KdV equation  \r{pkdv} for some function $\delta$.  

\smallskip

\noindent The reason for the phrase ``possibly with singularities''
here is because for some values of $x$ and $t$, $U(x,t,\l)$ might 
leave the dense open set of $LGL_2({\bf C})$ where Birkhoff decomposition 
is possible (it can be proved that these values are isolated). It is 
important for the purposes of this paper to note that 
although the above proposition makes no mention of the linear 
constant-coefficient flow \r{lin1}, the heart of its proof is that the  
this flow induces, via Birkhoff decomposition, 
the matrix Lax pair \r{lp1}-\r{lp2}
for the potential KdV equation. Note also that the first equation of the 
Lax pair \r{lp1} gives the usual relation of KdV with the Schr\"odinger 
equation. Writing either column of $Y$ as $\pmatrix{\psi \cr \phi \cr}$,
\r{lp1} gives $\psi_{xx}=(\l-2b_x)\psi$. 

There are many applications of the above result, of which only one will
be discussed here, the construction of the standard
B\"acklund transformation for potential
KdV. The idea behind this B\"acklund transformation is as follows:
Suppose the  element $U(0,0,\l)$ of the loop group gives solution 
$U(x,t,\l)$ of the linear system \r{lin1} with 
Birkhoff decomposition $S^{-1}Y$ and corresponding potential KdV solution
$b(x,t)$. Let us now try to find the potential KdV solution corresponding 
to the element
\be 
\sqrt{\frac{\l-\t}{\l}}
\pmatrix{ 0 & 1 \cr \l & 0 \cr} 
U(0,0,\l)
\pmatrix{ 0 & \frac1{\l-\t} \cr 1 & 0 \cr} 
\ee
with $0<\t<1$. The new solution of the linear system \r{lin1} is
\bea 
&&\sqrt{\frac{\l-\t}{\l}}
\pmatrix{ 0 & 1 \cr \l & 0 \cr} 
U(x,t,\l)
\pmatrix{ 0 & \frac1{\l-\t} \cr 1 & 0 \cr} \nonumber\\
&=&
\sqrt{\frac{\l-\t}{\l}}
\pmatrix{ 0 & 1 \cr \l & 0 \cr} 
S^{-1}(x,t,\l) Y(x,t,\l)
\pmatrix{ 0 & \frac1{\l-\t} \cr 1 & 0 \cr}\ . 
\eea
To perform the new Birkhoff decomposition, a certain matrix and 
its inverse must be inserted as follows:
\be
\sqrt{\frac{\l-\t}{\l}}
\pmatrix{ 0 & 1 \cr \l & 0 \cr} 
S^{-1}(x,t,\l) 
\pmatrix{ \alpha & 1 \cr \l-\t+\alpha\beta & \beta \cr}^{-1}
\cdot
\pmatrix{ \alpha & 1 \cr \l-\t+\alpha\beta & \beta \cr}
Y(x,t,\l)
\pmatrix{ 0 & \frac{1}{\l-\t} \cr 1 & 0 \cr} \ .
\ee
The aim is to chose $\alpha$ and $\beta$ so that this is written
in Birkhoff factorized form, i.e. so that 
\be 
\tilde{S}(x,t,\l) =
\sqrt{\frac{\l}{\l-\t}}
\pmatrix{ \alpha & 1 \cr \l-\t+\alpha\beta & \beta \cr}
S(x,t,\l) 
\pmatrix{ 0 & \frac{1}{\l} \cr 1 & 0 \cr} 
\la{scon}\ee
is holomorphic in $|\l|>1$ and satisfies $\tilde{S}(x,t,\infty) = I$, and 
\be
\tilde{Y}(x,t,\l) =
\pmatrix{ \alpha & 1 \cr \l-\t+\alpha\beta & \beta \cr}
Y(x,t,\l)
\pmatrix{ 0 & \frac{1}{\l-\t} \cr 1 & 0 \cr}  
\ee
is holomorphic in $|\l|<1$. Inserting the expansion \r{sexp} in
\r{scon}, the former condition requires $\beta=b$. For the 
latter condition it is just necessary to check $\tilde{Y}$ does not have a 
pole at $\l=\t$, and 
this requires $\alpha=-Y_{21}(x,t,\t)/Y_{11}(x,t,\t)$.
Finally, it is necessary to compute the new solution of the potential KdV
equation, i.e. the component of $1/\l$ in the 1,2-entry of $\tilde{S}$. 
A brief calculation shows this is simply $-\alpha$. Utilizing the Lax
pair \r{lp1}-\r{lp2} it is straightforward to determine properties 
of $\alpha$ leading to the following result:

\smallskip

\noindent{\bf Proposition 2.2:} If $b$ is a solution of the potential
KdV equation \r{pkdv} and $\psi$ satisfies 
\be
\psi_{xx} = (\t-2b_x) \psi\ ,\qquad
\psi_t = -{\textstyle{1\over 2}}b_{xx}\psi + (\t+b_x)\psi_x 
\la{lp}\ee
then $\tilde{b}=b+\psi_x/\psi$ is also a solution of potential KdV, for 
the same function $\delta$. 

\smallskip

Equations \r{lp} comprise the standard scalar Lax pair for the KdV
equation. Applying the B\"acklund transformation to the 
$x$-independent solution $b(t)=\int \delta(t)dt$ gives the 
1-soliton solutions
\be
b(x,t) = \int \delta(t) dt 
+ \sqrt{\t}{\rm tanh}\left(\sqrt{\t}(x+\t t)+C\right)
\ee
and the singular solutions
\be
b(x,t) = \int \delta(t) dt 
+ \sqrt{\t}{\rm coth}\left(\sqrt{\t}(x+\t t)+C\right)
\ee
where in both formulae $C$ is a constant.The easiest way to 
apply the B\"acklund 
transformation again to this solution is to use the {\em Bianchi 
permutability theorem} that states that the two-parameter 
family of solutions obtained by applying 
first the B\"acklund transformation with parameter $\t_1$  and then 
the B\"acklund transformation with parameter $\t_2$ is the same as the 
two-parameter family of solutions obtained by applying the two B\"acklund 
transformations in the reverse order. See \c{me1} for a detailed
discussion of this. The Bianchi 
permutability theorem can be used to derive 
an algebraic expression 
for the solutions obtained by applying two B\"acklund 
transformations (see \c{dj} section 5.4.3): 

\smallskip

\noindent{\bf Proposition 2.3:} If $b$ is a solution of the potential
KdV equation \r{pkdv}, and $b_1$ and $b_2$ are solutions 
obtained by applying  
B\"acklund transformations with parameters $\t_1$
and $\t_2$ respectively to $b$, then 
\be 
B = b + \frac{\t_1-\t_2}{b_1-b_2}
\ee
is a solution obtained by applying 
the two B\"acklund transformations successively 
to $b$, in either order. 

\smallskip

\noindent Applying this result  using a 1-soliton solution
for $b_1$, a singular solution for $b_2$ and $\t_2>\t_1$ gives 
the 2-soliton solution 
\bea
b(x,t) &=& \int \delta(t) dt +
\frac{\t_1-\t_2}{
\sqrt{\t_1}{\tanh}\alpha_1-
\sqrt{\t_2}{\coth}\alpha_2 
}  
\qquad
\left\{
\matrix{
\alpha_1=\sqrt{\t_1}(x+\t_1t)+C_1 \cr
\alpha_2=\sqrt{\t_2}(x+\t_2t)+C_2 \cr
}
\right.
\nonumber\\
&=& \int \delta(t) dt + \sqrt{\t_1}{\tanh}\alpha_1
                      + \sqrt{\t_2}{\tanh}\alpha_2 \\
&&~~~~~~~~~~~~~- \frac
{{\t_1}\tanh\alpha_2\ {\rm sech}^2 \alpha_1 + 
 \sqrt{\t_1\t_2}\tanh\alpha_1\ {\rm sech}^2 \alpha_2 }
{\sqrt{\t_2} - \sqrt{\t_1}\tanh\alpha_1\tanh\alpha_2  }
\nonumber 
\eea

This concludes our presentation of the basic theory of the KdV
equation and its relation with the linear system \r{lin1} which 
will be imitated for discrete systems in later sections.

\section{Discretizations I: Lattice KdV}

The aim in this section is to follow the procedures of the 
last section as closely as possible, but replacing the solution
$U(x,t,\l)$ of \r{lin1} by the solution $U_{nm}(\l)$ of the 
lattice equation \r{disc1}, which has general solution
\bea
U_{n,m}(\l) &=& \left[ I + h\pmatrix{ 0 & 1 \cr \l & 0 \cr} \right]^n
             \left[ I + k\pmatrix{ 0 & 1 \cr \l & 0 \cr} \right]^m 
   U_{0,0}(\l)  \nonumber\\
&=& \frac14
\pmatrix{ (1+h\sqrt{\l})^n+(1-h\sqrt{\l})^n   &
          \frac1{\sqrt{\l}}\left(  (1+h\sqrt{\l})^n-(1-h\sqrt{\l})^n \right)  \cr
          {\sqrt{\l}} \left( (1+h\sqrt{\l})^n-(1-h\sqrt{\l})^n \right)  &
          (1+h\sqrt{\l})^n+(1-h\sqrt{\l})^n   \cr } \la{huge} \\
&& \pmatrix{ (1+k\sqrt{\l})^m+(1-k\sqrt{\l})^m &
          \frac1{\sqrt{\l}} \left( (1+k\sqrt{\l})^m-(1-k\sqrt{\l})^m \right)  \cr
          {\sqrt{\l}}\left(  (1+k\sqrt{\l})^m-(1-k\sqrt{\l})^m \right)  &
          (1+k\sqrt{\l})^m+(1-k\sqrt{\l})^m   \cr }
U_{0,0}(\l)\ .  \nonumber
\eea  
Suppose $U_{n,m}(\l)$ has a Birkhoff factorization $S_{n,m}^{-1}(\l)Y_{n,m}(\l)$.  
Substituting in \r{disc1} and rearranging gives
\bea
Y_{n+1,m}Y_{n,m}^{-1} &=& S_{n+1,m}
\left[ I + h\pmatrix{ 0 & 1 \cr \l & 0 \cr} \right] S_{n,m}^{-1}\ , \la{int1}\\
Y_{n,m+1}Y_{n,m}^{-1} &=& S_{n,m+1}
\left[ I + k\pmatrix{ 0 & 1 \cr \l & 0 \cr} \right] S_{n,m}^{-1}\ . \la{int2}
\eea
Writing  
\be S_{n,m} = 
I + \frac1{\l} \pmatrix{ a_{n,m} & -b_{n,m} \cr
                           c_{n,m} & d_{n,m} \cr } + \ldots
\ee
and comparing non-negative powers of $\l$ on both sides of \r{int1}-\r{int2}
gives 
\bea
Y_{n+1,m} &=& 
\pmatrix{ 1-hb_{n+1,m}  & h \cr h\l + h(d_{n+1,m}-a_{n,m}) & 1+hb_{n,m} \cr } Y_{n,m}\ ,
\la{int3}\\
Y_{n,m+1}  &=& 
\pmatrix{ 1-kb_{n,m+1}  & k \cr k\l + k(d_{n,m+1}-a_{n,m}) & 1+kb_{n,m} \cr } Y_{n,m}\ .
\la{int4}
\eea
There is one further simplification that can be made in these equations.  
\r{int1} (and similarly \r{int2}) can be written in the form
\be
S_{n,m}S_{n+1,m}^{-1} \cdot  Y_{n+1,m}Y_{n,m}^{-1} =  
S_{n,m}\left[ I + h\pmatrix{ 0 & 1 \cr \l & 0 \cr} \right] S_{n,m}^{-1}\ .
\ee
Taking the determinant gives
\be 
\det\left( S_{n,m}S_{n+1,m}^{-1} \right)
\det\left( Y_{n+1,m}Y_{n,m}^{-1} \right) =
1-h^2\l \ . 
\ee
The Birkhoff factorization theorem applies for scalars (elements of $LGL_1{\bf C}$) too, 
so from this it can be deduced that $\det\left(  Y_{n+1,m}Y_{n,m}^{-1} \right) = 1-h^2\l$ 
(and $\det\left( S_{n,m}S_{n+1,m}^{-1} \right)=1$). Applying this to \r{int3} (and the 
corresponding result $\det\left(  Y_{n,m+1}Y_{n,m}^{-1} \right) = 1-k^2\l$  to \r{int4})
gives 
\bea
Y_{n+1,m} &=& 
\pmatrix{ 1-hb_{n+1,m}  & h \cr h\l +
b_{n,m}-b_{n+1,m}-hb_{n,m}b_{n+1,m}
 & 1+hb_{n,m} \cr } Y_{n,m}\ ,
\la{dlp1}\\
Y_{n,m+1}  &=& 
\pmatrix{ 1-kb_{n,m+1}  & k \cr k\l + 
b_{n,m}-b_{n,m+1}-kb_{n,m}b_{n,m+1}
& 1+kb_{n,m} \cr } Y_{n,m}\ .
\la{dlp2}
\eea
Up to a rescaling this is precisely Nijhoff {\em et al.}'s scalar Lax pair for 
the lattice KdV equation \c{Nijhoff}. Writing
\bea   
L_{n,m} &=& \pmatrix{ 1-hb_{n+1,m}  & h \cr h\l +
b_{n,m}-b_{n+1,m}-hb_{n,m}b_{n+1,m}
 & 1+hb_{n,m} \cr } \ , \\
M_{n,m} &=& \pmatrix{ 1-kb_{n,m+1}  & k \cr k\l + 
b_{n,m}-b_{n,m+1}-kb_{n,m}b_{n,m+1}
& 1+kb_{n,m} \cr } \ ,
\eea
equations \r{dlp1}-\r{dlp2} are just 
\be    
Y_{n+1,m}=L_{n,m}Y_{n,m}\ , \qquad 
Y_{n,m+1}=M_{n,m}Y_{n,m}\ ,
\la{cons}\ee
and for consistency $L_{n,m+1}M_{n,m}=M_{n+1,m}L_{n,m}$.
(This last equation plays the role of the zero-curvature equation in 
the continuous case.) Substituting the forms found for 
$L_{n,m},M_{n,m}$ in the consistency condition gives
lattice KdV \r{dp2}. Thus the analog of Proposition 2.1 is obtained:

\smallskip

\noindent{\bf Proposition 3.1:} 
Let $U_{0,0}(\l)$ be an element of $LGL_2({\bf C})$, 
let $S_{n,m}^{-1}(\l)Y_{n,m}(\l)$ be the Birkhoff decomposition of 
\be
U_{n,m}(\l) = \left[ I + h\pmatrix{ 0 & 1 \cr \l & 0 \cr} \right]^n
              \left[ I + k\pmatrix{ 0 & 1 \cr \l & 0 \cr} \right]^m 
   U_{0,0}(\l)\ ,
\ee
and let 
$b_{n,m}$ be $(-1)$ times the component of $1/\l$ in the 1,2-entry of 
$S_{n,m}$. Then $b_{n,m}$ is a solution, possibly with singularities, of
the lattice KdV equation  \r{dp2}.  

\smallskip

\noindent In fact there is no reason why $U_{0,0}(\l)$ should not, in this
case, be dependent on $h$ and $k$. So in principle the class of solutions 
of lattice KdV occuring this way is much larger than the corresponding class of 
solutions of (potential) KdV. 

Let us attempt to find a B\"acklund transformation and soliton solutions for 
lattice KdV proceeding as in section 2. 
Making the replacement 
\be 
U_{0,0}(\l) \rightarrow 
\sqrt{\frac{\l-\t}{\l}}
\pmatrix{ 0 & 1 \cr \l & 0 \cr} 
U_{0,0}(\l)
\pmatrix{ 0 & \frac1{\l-\t} \cr 1 & 0 \cr} 
\ee
gives 
\be 
U_{n,m}(\l) \rightarrow 
\sqrt{\frac{\l-\t}{\l}}
\pmatrix{ 0 & 1 \cr \l & 0 \cr} 
U_{n,m}(\l)
\pmatrix{ 0 & \frac1{\l-\t} \cr 1 & 0 \cr} 
\ee
and 
\bea 
S_{n,m}(\l) &\rightarrow&
\sqrt{\frac{\l}{\l-\t}}
\pmatrix{ \alpha_{n,m}
 & 1 \cr \l-\t+\alpha_{n,m}b_{n,m} &b_{n,m} \cr}
S_{n,m}(\l) 
\pmatrix{ 0 & \frac{1}{\l} \cr 1 & 0 \cr} 
\\
Y_{n,m}(\l) &\rightarrow&
\pmatrix{ \alpha_{n,m}
 & 1 \cr \l-\t+\alpha_{n,m}b_{n,m} & b_{n,m} \cr}
Y_{n,m}(\l)
\pmatrix{ 0 & \frac{1}{\l-\t} \cr 1 & 0 \cr}  
\eea
where $\alpha_{n,m}=-(Y_{n,m})_{21}(\t)/(Y_{n,m})_{11}(\t)$.
The new solution of lattice KdV is simply $-\alpha_{n,m}$. Using
\r{dlp1}-\r{dlp2} to find properties of $\alpha_{n,m}$ gives
the B\"acklund transformation:

\smallskip

\noindent{\bf Proposition 3.2:} If $b_{n,m}$ is a solution of the 
lattice KdV equation \r{dp2} and $\psi_{n,m}$ satisfies 
\bea
\frac{\psi_{n+2,m}-2\psi_{n+1,m}+\psi_{n,m}}{h^2}
&=& \t\psi_{n,m}- \left(\frac{b_{n+2,m}-b_{n,m}}{h}\right) \psi_{n+1,m} 
\la{dsl1}\\
\frac{\psi_{n,m+1}-\psi_{n,m}}{k}
&=& \frac{\psi_{n+1,m}-\psi_{n,m}}{h} + (b_{n+1,m}-b_{n,m+1})\psi_{n,m}
\la{dsl2}
\eea
then $\tilde{b}_{n,m}=b_{n+1,m}+(\psi_{n+1,m}-\psi_{n,m})/(h\psi_{n,m})$ 
is also a solution of lattice KdV. 

\smallskip

\noindent The first equation here is a natural discretization of the 
first equation in  \r{lp}, and is the discretization of the Schr\"odinger 
equation studied in \c{BPPS}. 
The second equation is, however, completely
unrelated to that in \r{lp}. To get 1-soliton and singular solutions
the B\"acklund transformation can be applied to the trivial solution 
$b_{n,m}=0$. This gives solutions of the form
\be
b_{n,m}= \sqrt{\t} 
\frac{ A(1+h\sqrt{\t})^n(1+k\sqrt{\t})^m -  B(1-h\sqrt{\t})^n(1-k\sqrt{\t})^m }
     { A(1+h\sqrt{\t})^n(1+k\sqrt{\t})^m +  B(1-h\sqrt{\t})^n(1-k\sqrt{\t})^m }
\qquad (A,B~{\rm constants}).
\la{gensol}
\ee
If $A:B$ is positive and $h,k<1/\sqrt{\t}$ this gives 1-soliton solutions 
\be
b_{n,m}=\sqrt{\t}\tanh\left(
n \tanh^{-1}(h\sqrt{\t}) + 
m \tanh^{-1}(k\sqrt{\t}) + C
\right)\ 
\qquad (C~{\rm constant}).
\la{dsol}\ee
If $A:B$ is negative and $h,k<1/\sqrt{\t}$ \r{gensol} gives singular solutions 
\be
b_{n,m}=\sqrt{\t}\coth\left(
n \tanh^{-1}(h\sqrt{\t}) + 
m \tanh^{-1}(k\sqrt{\t}) + C
\right)\ .
\ee

% Important identity:  
% \frac12  \ln\left( \frac{1+h\sqrt{\t}}{1-h\sqrt{\t}} \right) 
%  =  \tanh^{-1} (h\sqrt{\t}) 

The Bianchi permutability theorem applies equally here in the 
discrete case, and this can be used to give the discrete version of
Proposition 2.3:

\smallskip

\noindent{\bf Proposition 3.3:} If $b_{n,m}$ is a solution of the lattice
KdV equation \r{dp2}, and $b^{(1)}_{n,m}$ and $b^{(2)}_{n,m}$ are solutions 
obtained by applying
B\"acklund transformations with parameters $\t_1$
and $\t_2$ respectively to $b_{n,m}$, then 
\be 
B_{n,m} = b_{n,m}
 + \frac{\t_1-\t_2}{b^{(1)}_{n,m}-b^{(2)}_{n,m}}
\ee
is a solution obtained by applying 
the two B\"acklund transformations successively 
to $b_{n,m}$, in either order. 

\smallskip

\noindent{\bf Proof.} Writing $q_{n,m}=(\psi_{n+1,m}-\psi_{n,m})/(h\psi_{n,m})$,
the B\"acklund transformation can be written as $b_{n,m}\rightarrow 
b_{n+1,m}+q_{n,m}$ where $q_{n,m}$ satisifies the discrete Riccati equation \c{ss}
\be
q_{n+1,m} = \frac{q_{n,m}(1-hb_{n+2,m}+hb_{n,m}) + (h\t-b_{n+2,m}+b_{n,m})}
                 {1+hq_{n,m}} \ 
\ee
(for the sake of brevity I only look at the first equation in \r{dsl1}-\r{dsl2}).
Alternatively, after a little algebra, the transformation can be written
$b_{n,m}\rightarrow \tilde{b}_{n,m}$
where $b_{n,m}$, $\tilde{b}_{n,m}$ are related by
\be
\frac{\tilde{b}_{n+1,m} - \tilde{b}_{n,m} + b_{n+1,m} -b_{n,m}}{h}
=\t + (b_{n,m}-\tilde{b}_{n+1,m})(\tilde{b}_{n,m}-b_{n+1,m})\ .
\ee
Using the Bianchi permutability theorem and the premises of the theorem, it is
known that
applying the BT with parameter $\t_1$ to $b_{n,m}$ gives $b^{(1)}_{n,m}$, 
applying the BT with parameter $\t_2$ to $b_{n,m}$ gives $b^{(2)}_{n,m}$, 
and
applying {\em either} the BT with parameter $\t_2$ to $b^{(1)}_{n,m}$ 
{\em or} the BT with parameter $\t_1$ to $b^{(2)}_{n,m}$ gives 
the same solution $B_{n,m}$.
This implies 4 equations:
\bea
\frac{b^{(1)}_{n+1,m} - b^{(1)}_{n,m} + b_{n+1,m} -b_{n,m}}{h}
&=&\t_1 + (b_{n,m}-b^{(1)}_{n+1,m})(b^{(1)}_{n,m}-b_{n+1,m}) \ ,    \\
\frac{b^{(2)}_{n+1,m} - b^{(2)}_{n,m} + b_{n+1,m} -b_{n,m}}{h}
&=&\t_2 + (b_{n,m}-b^{(2)}_{n+1,m})(b^{(2)}_{n,m}-b_{n+1,m}) \ , \\
\frac{B_{n+1,m} - B_{n,m} + b^{(1)}_{n+1,m} -b^{(1)}_{n,m}}{h}
&=&\t_2 + (b^{(1)}_{n,m}-B_{n+1,m})(B_{n,m}-b^{(1)}_{n+1,m}) \ , \\ 
\frac{B_{n+1,m} - B_{n,m} + b^{(2)}_{n+1,m} -b^{(2)}_{n,m}}{h}  
&=&\t_1 + (b^{(2)}_{n,m}-B_{n+1,m})(B_{n,m}-b^{(2)}_{n+1,m})   \ .
\eea
\smallskip
Adding the first and last of these equations and subtracting the other two gives
\be
2(\t_1-\t_2) = (B_{n,m} - b_{n,m} )(b^{(1)}_{n,m} - b^{(2)}_{n,m})
+ (B_{n+1,m} - b_{n+1,m} )(b^{(1)}_{n+1,m} - b^{(2)}_{n+1,m})
\ee
The general solution of this is clearly
\be 
(B_{n,m} - b_{n,m} )(b^{(1)}_{n,m} - b^{(2)}_{n,m} )
= (\t_1-\t_2) + (-1)^n F(m)\ ,
\ee
where $F$ is an arbitrary function of $m$. Using the second equation in
\r{dsl1}-\r{dsl2} is is possible to show that $F(m)=0$. $\bullet$. 

All that remains to do in this section is to briefly discuss 
the nature of soliton solutions of lattice KdV,  and in particular
how they compare to those of continuum KdV. From \r{dsol} the 
speed of the soliton with parameter $\t$ is 
\be 
c=\frac{h\tanh^{-1}(k\sqrt{\t})}{k\tanh^{-1}(h\sqrt{\t})}\ . 
\ee
(The formal definition of the ``speed'' is the number $c$ such that 
the solution depends on $m,n$ only through the combination $nh+cmk$.)
Recall that the parameter $\t$ is limited by the requirements
$h\sqrt{\t},k\sqrt{\t}<1$. Thus:

\smallskip

\noindent{\bf Proposition 3.4:} 
For $h=k$ the soliton solutions of lattice KdV \r{dp2} all have speed $1$. 
For $h<k$ there are solitons with all speeds greater than $1$. 
For $h>k$ there are solitons with all speeds between $0$ and  $1$. 

\noindent{\bf Proof.} The result for $h=k$ is obvious. Switching $h$ and $k$ 
switches $c$ and $1/c$ so  it is  just necessary to check the result for, say, $h<k$.
As $\t$ tends to $0$ $c$ tends to $1$, and as $\t$ tends to $1/k^2$ (which is 
less than $1/h^2$) $c$ tends to $\infty$. So the result will be proved if we 
can establish that $c$ is a monotonic increasing function 
of $\t$ for $0<\t<1/k^2$. 
Writing $z=k\sqrt{\t}$ and $\alpha=h/k<1$, 
\be c = \frac{\alpha \tanh^{-1} z}{\tanh^{-1} \alpha z}\ ,  \ee
and it is necessary
to check this is a monotonic function of $z$ on $0<z<1$ for 
$\alpha$ fixed between $0$ and $1$. Differentiating gives 
\be \frac{dc}{dz} =  
\frac{\alpha}
{[\tanh^{-1}(\alpha z)]^2(1-z^2)(1-\alpha^2 z^2)}
\left[ (1-\alpha^2 z^2)\tanh^{-1}(\alpha z) - \alpha(1-z^2)\tanh^{-1}(z)
\right]\ .
\ee
All the terms except the last are evidently positive. The last term 
can be written $g(\alpha z) - \alpha g(z)$  where $g(z)=(1-z^2)\tanh^{-1}(z)$.
Thus it is necessary to show $g(\alpha z)> \alpha g(z)$. But this follows 
immediately from the convexity of $g$, which is trivial as 
\be \frac{d^2g}{dz^2} = -2 \tanh^{-1} z - \frac{2z}{1-z^2} < 0 
\qquad {\rm for~}0<z<1\ . \qquad \bullet \ee

\smallskip

Proposition 3.4  does a lot to clarify the relationship of lattice KdV
\r{dp2} with its standard continuum limit $b_t=b_x$ on the one hand, 
and potential KdV \r{pkdv} on the other. The linear equation 
$b_t=b_x$ admits solitons  of speed $1$, but, since it is linear, 
the solitons can be of arbitrary amplitude. The indirect method 
of discretization used
has given rise to a {\em nonlinear} discretization, except when 
$k=h$ (when \r{dp2} can be written as a product of linear factors). 
The family of speed $1$ solitons with arbitrary amplitude is perturbed,
after discretization, into a family of solitons with a nontrivial 
speed-amplitude relation. For small $h,k$ the low amplitude solitons 
(those with $\sqrt{\t}\ll 1/h,1/k$) must have speed close to $1$, and indeed 
this is the case. For larger amplitudes the speeds can change substantially, 
giving a range of speeds ranging from $1$ to either $0$ or $\infty$. 
Since there now are solitons of different speeds, and the 
necessary algebraic structure has been preserved, the phenomena associated with KdV
will emerge, in particular elastic soliton scattering. Thus from a phenomenological 
viewpoint, lattice KdV is closer to potential KdV than the linear
equation $b_t=b_x$. There are, however,  several fundamental differences:
First, the range of soliton speeds in lattice KdV is limited to speeds either 
less than or greater than $1$. Second, there are many solutions of lattice
KdV that do not have natural continuum limits; for example, solutions 
\r{gensol} in the case where $\t$ exceeds  $1/h$ or $1/k$ (or both). 

\section{Discretizations II: The simplest natural discretization}

This section is devoted  to the simple discretization \r{disc2} of
\r{lin1}, which, as explained in the introduction, should give an
integrable lattice equation which has potential KdV as its 
{\em standard} continuum limit. The general solution of \r{disc2}
is given by \r{huge} on replacing $k$ with $k\l$. 

Once again 
suppose $U_{n,m}(\l)$ has a Birkhoff factorization $S_{n,m}^{-1}(\l)Y_{n,m}(\l)$,
and substitute in \r{disc2} to get 
\bea
Y_{n+1,m}Y_{n,m}^{-1} &=& S_{n+1,m}
\left[ I + h\pmatrix{ 0 & 1 \cr \l & 0 \cr} \right] S_{n,m}^{-1}\ , \la{tm1}\\
Y_{n,m+1}Y_{n,m}^{-1} &=& S_{n,m+1}
\left[ I + k\l\pmatrix{ 0 & 1 \cr \l & 0 \cr} \right] S_{n,m}^{-1}\ . 
\la{tm2}\eea
Writing 
\be S_{n,m} = 
I + \frac1{\l} \pmatrix{ a_{n,m} & -b_{n,m} \cr
                           c_{n,m} & d_{n,m} \cr } 
+ \frac1{\l^2} \pmatrix{ \tilde{a}_{n,m} & \tilde{b}_{n,m} \cr
                           \tilde{c}_{n,m} & \tilde{d}_{n,m} \cr } + \ldots\ ,
\la{S}\ee
and employing the relations $\det\left(Y_{n+1,m}Y_{n,m}^{-1} \right)=1-h^2\l$, 
$\det\left(Y_{n,m+1}Y_{n,m}^{-1} \right)=1-k^2\l^3$ (obtained by left-multiplying 
\r{tm1} and \r{tm2} respectively by $S_{n,m}S_{n+1,m}^{-1}$  and
$S_{n,m}S_{n,m+1}^{-1}$, taking the determinant and factorizing) gives
\be  
Y_{n+1,m}=L_{n,m}Y_{n,m}\ , \qquad 
Y_{n,m+1}=M_{n,m}Y_{n,m}\ ,
\ee
where
\bea
L_{n,m}&=&
\pmatrix{ 1-hb_{n+1,m}  & h \cr h\l +
b_{n,m}-b_{n+1,m}-hb_{n,m}b_{n+1,m}
 & 1+hb_{n,m} \cr } \ , \\
M_{n,m}&=& k
\pmatrix{
- \lambda b_{n,m+1} +  \Sigma_{n,m}+\Delta_{n,m}-\beta_{n,m} b_{n,m+1} &
\lambda + \beta_{n,m} \cr
\cr
\matrix{
\lambda^2 - \lambda(\beta_{n,m}+b_{n,m}b_{n,m+1}) 
-b_{n,m}b_{n,m+1}\beta_{n,m}  + \cr
\beta_{n,m}^2 +\D_{n,m}(b_{n,m}+b_{n,m+1})
+\Sigma_{n,m}(b_{n,m}-b_{n,m+1}) \cr}  &
\matrix{
\lambda b_{n,m} +  \Sigma_{n,m} - \cr 
\D_{n,m} +\b_{n,m} b_{n,m}\cr}  \cr
} \ .
\eea

\smallskip
\noindent The matrix $M$ depends on three lattice fields $\b,\D,\Sigma$ in addition 
to the basic lattice field $b$, but $\Sigma$ is determined via the relation
\be
\Sigma_{n,m}
= \sqrt{\frac1{k^2} + \Delta_{n,m}^2 + \beta_{n,m}^3}\ .
\ee
Substituting these ans\"atze into the consistency equation \r{cons} gives
the following 3 equations for the 3 fundamental fields $b,\b,\D$:
\be
\b_{n+1,m} + \b_{n,m} =
   \frac{b_{n+1,m}+b_{n+1,m+1}-b_{n,m}-b_{n,m+1} }{h}
  + (b_{n+1,m+1}-b_{nm})(b_{n,m+1}-b_{n+1,m})  \ ,
\la{fkdv1}\ee
\be
\D_{n+1,m} + \D_{n,m} =
   \left(\frac{\b_{n+1,m}-\b_{n,m}  }{h}\right)
   \left(-1 + \frac{h}2(b_{n+1,m+1}+b_{n+1,m}-b_{n,m+1}-b_{n,m}) \right) \ ,
\la{fkdv2}\ee
\bea
 \left( \frac{\b_{n+1,m}-\b_{n,m} }{h} \right)
 \left( \frac{b_{n,m+1}+b_{n+1,m+1}-b_{n,m}-b_{n+1,m}}{k} \right)
~~~~~~~~~~~~~~~~~~\nonumber  \\
~~~~~~~~~~~~~~~~~~=
\frac{
\sqrt{1+k^2(\D_{n+1,m}^2+\b_{n+1,m}^3)} 
- \sqrt{1+k^2(\D_{n,m}^2+\b_{n,m}^3)} 
}{{\textstyle{\frac12}}hk^2} \ .
\la{fkdv3}\eea
Note the equations involve $b$ at 4 points 
($b_{n,m}$, $b_{n+1,m}$, $b_{n,m+1}$, $b_{n+1,m+1}$) but $\b$ and $\D$ at only 2
($\b_{n,m}$,$\b_{n+1,m}$,$\D_{n,m}$,$\D_{n+1,m}$). 
The system \r{fkdv1}-\r{fkdv3} will be given the title {\em full lattice KdV};
as will shortly be shown, unlike standard lattice KdV,  full lattice KdV
displays, for certain choices of $h$ and $k$,
solitons with the full range of speeds. Full lattice KdV also has, as
expected, potential KdV as a standard continuum limit: Replacing 
$b_{n,m}$ by $b(x,t)$, $b_{n+1,m}$ by $b(x+h,t)$, $b_{n,m+1}$ by $b(x,t+k)$, 
$b_{n+1,m+1}$ by $b(x+h,t+k)$, and similarly for $\b$ and $\D$, and then 
taking the limit $h,k\rightarrow 0$, the equations \r{fkdv1}-\r{fkdv3} become 
\be 
2\b = 2b_x\ ,  \qquad
2\D = -\b_x\ ,  \qquad
2\b_x b_t =  (\D^2+\b^3)_x \ .
\ee
Eliminating $\b$ and $\D$ from these yields potential KdV
$b_t = \frac14 b_{xxx} + \frac32 b_x^2$\ .

There are analogs for full lattice KdV
of all the results of the previous sections:

\smallskip

\noindent{\bf Proposition 4.1:} 
Let $U_{0,0}(\l)$ be an element of $LGL_2({\bf C})$, 
let $S_{n,m}^{-1}(\l)Y_{n,m}(\l)$ be the Birkhoff decomposition of 
\be
U_{n,m}(\l) = \left[ I + h\pmatrix{ 0 & 1 \cr \l & 0 \cr} \right]^n
              \left[ I + k\l \pmatrix{ 0 & 1 \cr \l & 0 \cr} \right]^m 
   U_{0,0}(\l)\ ,
\ee
and let 
$b_{n,m}$ be $(-1)$ times the component of $1/\l$ in the 1,2-entry of 
$S_{n,m}$. Then $b_{n,m}$ is a solution, possibly with singularities, of
the full lattice KdV system  \r{fkdv1}-\r{fkdv3}. 

\smallskip

\noindent By ``$b_{n,m}$ is a solution of full lattice KdV,''
I mean that there exist fields $\b,\D$ for which equations \r{fkdv1}-\r{fkdv3}
hold. In practice, once $b$ is known, the easiest way to determine $\b,\D$ 
will be directly from equations \r{fkdv1} and \r{fkdv2}. In the previous 
proposition the other fields can actually be determined from $S$
if this is known in full: If the expansion of 
$S$ in powers of $1/\l$ is as in \r{S}, then 
\bea
\b_{n,m} &=&   a_{n,m+1}-d_{n,m}-b_{n,m}b_{n,m+1} \\
\D_{n,m} &=& \frac12\left( \tilde{b}_{n,m+1} +  \tilde{b}_{n,m}  - c_{n,m+1} - c_{n,m}  
            - b_{n,m}d_{n,m+1} - b_{n,m+1}d_{n,m}  \right.\nonumber \\
  &&    \left.+ (b_{n,m+1}+b_{n,m}) (a_{n,m}+a_{n,m+1}- b_{n,m}b_{n,m+1})
             \right)  \\  
\Sigma_{n,m} &=& \frac1{k}+\frac12
        \left( \tilde{b}_{n,m+1}  -  \tilde{b}_{n,m}  + c_{n,m+1} - c_{n,m} 
            + b_{n,m}d_{n,m+1} - b_{n,m+1}d_{n,m}  \nonumber \right.\\
  &&  \left.  + (b_{n,m+1}-b_{n,m}) (a_{n,m}+a_{n,m+1}- b_{n,m}b_{n,m+1})  \right)
\eea  
The B\"acklund transformation takes the following form:

\smallskip

\noindent{\bf Proposition 4.2:} If $b_{n,m},\b_{n,m},\D_{n,m}$ is a solution of the 
full lattice KdV equation \r{fkdv1}-\r{fkdv3} and $\psi_{n,m}$ satisfies 
\bea
\frac{\psi_{n+2,m}-2\psi_{n+1,m}+\psi_{n,m}}{h^2}
&=& \t\psi_{n,m}- \left(\frac{b_{n+2,m}-b_{n,m}}{h}\right) \psi_{n+1,m} 
\la{fsl1}\\
\frac{\psi_{n,m+1}-\psi_{n,m}}{k}
&=& (\t+\b_{n,m})\frac{\psi_{n+1,m}-\psi_{n,m}}{h}   \la{fsl2}\\
&&+ \left( (b_{n+1,m}-b_{n,m+1})(\t+\b_{n,m}) + \D_{n,m} + \Sigma_{n,m}-\frac1{k}   
  \right) \psi_{n,m}\nonumber 
\eea
then $b^{\rm new}_{n,m}=b_{n+1,m}+(\psi_{n+1,m}-\psi_{n,m})/(h\psi_{n,m})$ 
is also a solution of full lattice KdV.  The fields $\b,\D$ are replaced by
$\b^{\rm new},\D^{\rm new}$ respectively, which are given by the following algebraic
equations: 
\bea
\b^{\rm new}_{n,m} + \b_{n,m} &=& \t
+ (b_{n,m}-b^{\rm new}_{n,m+1})(b^{\rm new}_{n,m}-b_{n,m+1})\ ,  \la{alg1}\\
\D^{\rm new}_{n,m} + \D_{n,m} &=& \frac12
\left( \b^{\rm new}_{n,m} - \b_{n,m} \right)
\left( b^{\rm new}_{n,m+1} + b^{\rm new}_{n,m} - b_{n,m+1} - b_{n,m}  \right)\ .
\la{alg2}
\eea

\smallskip

\noindent The B\"acklund transformation for continuum KdV (Proposition 
2.2) is recovered in the limit $h,k\rightarrow 0$, since 
$\b_{n,m}\rightarrow b_x$, $\D_{n,m}\rightarrow -\frac12b_{xx}$, 
$\Sigma_{n,m}-\frac1{k}\rightarrow 0$. 
Note the difference between the second equation in \r{fsl1}-\r{fsl2}
and the discrete evolution proposed in \c{BPPS}. 
The solutions obtained 
using the B\"acklund transformation on the vacuum solution
$b_{n,m}=\b_{n,m}=\D_{n,m}=0$ are given by \r{gensol} with $k$ 
replaced by $k\t$. In particular, writing 
$t(n,m)$ in place of $\tanh\left(n \tanh^{-1}(h\sqrt{\t}) + 
m \tanh^{-1}(k\t\sqrt{\t}) + C\right)$, it is  
straightforward to verify  that the soliton solution is given by
\bea
b_{n,m}&=&\sqrt{\t}t(n,m)\ , \la{fsol}\\
\b_{n,m}&=&\frac{t(n,m+1)-t(n,m)}{k\sqrt{\t}}  \ ,\\
\D_{n,m}&=&\frac{t(n,m+1)^2-t(n,m)^2}{2k}  \ ,\\
\Sigma_{n,m} &=&\frac1{k}+  \frac{(t(n,m+1)-t(n,m))^2}{2k}\ . 
\eea
Before exploring the phenomenology of these solitons, note that since 
the proof of Proposition 3.3 is  based almost entirely on the first 
equation of the scalar Lax pair \r{dsl1}-\r{dsl2}, it is no surprise
that it goes through verbatim to full lattice KdV, i.e. 

\smallskip

\noindent{\bf Proposition 4.3:} If $b_{n,m}$ is a solution of the full lattice
KdV system \r{fkdv1}-\r{fkdv3}, and $b^{(1)}_{n,m}$ and $b^{(2)}_{n,m}$ 
are solutions obtained by applying
B\"acklund transformations with parameters $\t_1$
and $\t_2$ respectively to $b_{n,m}$, then 
\be 
B_{n,m} = b_{n,m}
 + \frac{\t_1-\t_2}{b^{(1)}_{n,m}-b^{(2)}_{n,m}}
\ee
is a solution obtained by applying 
the two B\"acklund transformations successively 
to $b_{n,m}$, in either order. 

\smallskip

\noindent Since the formulae \r{alg1}-\r{alg2} for applying the B\"acklund 
transformation to the fields $\b,\D$ are already pure algebraic there is 
no need to consider them in proposition 4.3. 

It just remains to investigate the speed-amplitude relation
of the soliton solutions. The soliton speed is
\be 
c=\frac{h\tanh^{-1}(k\t\sqrt{\t})}{k\tanh^{-1}(h\sqrt{\t})}\ ,
\ee
where the range of the parameter $\t$ is limited by the requirements 
$k\t\sqrt{\t},h\sqrt{\t}<1$. Writing $\alpha=hk^{-1/3}$ and 
$z=\sqrt{\t}k^{1/3}$ gives
\be c = k^{-2/3} \frac{\alpha \tanh^{-1}(z^3)}{\tanh^{-1}(\alpha z)}\ . \ee
See figure 1. For $\alpha< 1 \Leftrightarrow h^3 < k $ the speed is 
a monotonic increasing function of $z$ (or $\t$), going from $0$ as 
$z\rightarrow 0$ to $\infty$ as $z\rightarrow 1$. Thus for this range of 
parameters the soliton content exactly mirrors that of continuum KdV. 
For $\alpha= 1 \Leftrightarrow h^3 = k $ the speed is
a monotonic increasing function of $z$ (or $\t$), going from $0$ as 
$z\rightarrow 0$ to $1$ as $z\rightarrow 1$. 
For $\alpha> 1 \Leftrightarrow h^3 > k $ there is  an interesting effect 
that $c$ increases from $0$, reaches a maximum value, 
and then decreases again to $0$ as $z$ approaches $1/\alpha$.  
Thus for these choices of $h,k$ there is a limited set of speeds, but for 
all but the fastest there are solitons of $2$ different amplitudes; furthermore
these can be superposed to give other types of soliton solution. 
Note that 
if our interest in discretizations of KdV were for the purposes of numerical
simulation, we would presumably want both $h$ and $k$ small and of the same 
order of magnitude, and thus be in the $h^3<k$ regime, where the 
soliton phenomenology is correct.   

\begin{figure}
\centerline{\includegraphics{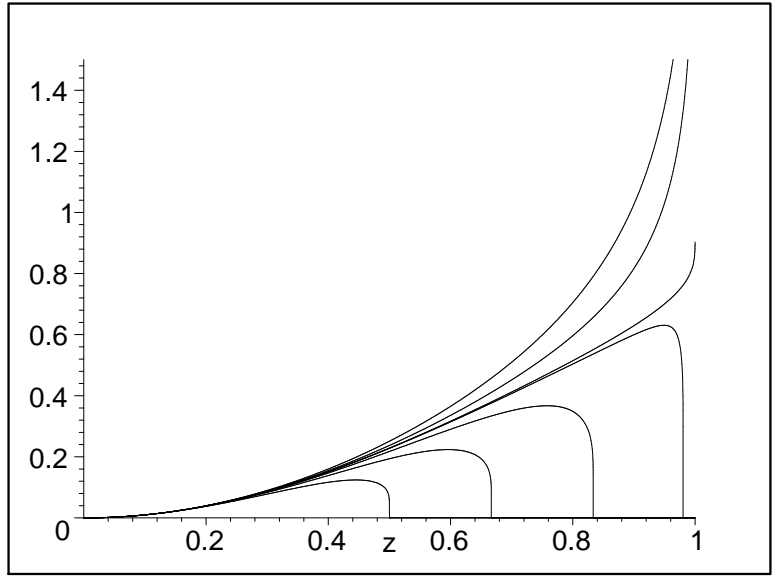}}
\caption{The function $\alpha\tanh^{-1}(z^3)/\tanh^{-1}(\alpha z)$ for 
$\alpha=0.1,0.8,1,1.02,1.2,1.5,2.0$ (from top to bottom)}
\end{figure}

\section{Discretizations III: A second-order discretization}

In this section our method is applied to the discretization
\r{so1}-\r{so2} of \r{lin2} with $N=1$. The resulting 
system is of limited intrinsic interest, the main point here 
is to illustrate that our methods can in principle be extended to give 
a whole range of integrable discretizations of equations in the KdV hierarchy.
One interesting point that emerges is the form of the related 
discretization of the Schr\"odinger equation. 

Following the usual procedure, assuming $U_{n,m}(\l)$ has a
Birkhoff decomposition $S_{n,m}^{-1}(\l)Y_{n,m}(\l)$, writing 
\be S_{n,m} = 
I + \frac1{\l} \pmatrix{ a_{n,m} & -b_{n,m} \cr
                           c_{n,m} & d_{n,m} \cr } + \ldots
\ee
etc., gives the system 
\be    
Y_{n+1,m}=L_{n,m}Y_{n,m}\ , \qquad 
Y_{n,m+1}=M_{n,m}Y_{n,m}\ ,
\ee
where 
\be
L_{n,m} = \pmatrix{ 1+\frac12 h^2\l-hb_{n+1,m}+ \frac12 h^2(a_{n+1,m}-a_{n,m}) & 
         h  +\frac12 h^2(b_{n,m}-b_{n+1,m}) \cr 
         h\l + b_{n,m}-b_{n+1,m}-h\D &
         1+ \frac12 h^2\l +  hb_{n,m}  + \frac12 h^2(a_{n,m}-a_{n+1,m})\cr } \ ,
\ee
\be
M_{n,m} = \pmatrix{ 1-kb_{n,m+1}  & 
         k  \cr 
         k\l + b_{n,m}-b_{n,m+1}-kb_{n,m}b_{n,m+1}  &
         1+   kb_{n,m}  \cr }\ ,
\ee
and 
\be
\D= 
\frac{\frac12(b_{n,m}^2+b_{n+1,m}^2)-\frac12h(a_{n+1,m}-a_{n,m})(b_{n+1,m}+b_{n,m})
+\frac14h^2(a_{n,m}-a_{n+1,m})^2}
{1+\frac12 h(b_{n,m}-b_{n+1,m})}  
\ee
The consistency condition $L_{n,m+1}M_{n,m}=M_{n+1,m}L_{n,m}$ 
unravels to two equations for the fields $a,b$.
Introducing the combinations 
\bea
\Sigma_{n,m}&=&\frac12\left(
a_{n+1,m+1} - a_{n,m+1} + a_{n+1,m} - a_{n,m}
\right)\ , \\
\Delta_{n,m}&=&\frac12\left(
a_{n+1,m+1} - a_{n,m+1} - a_{n+1,m} + a_{n,m}
\right)\ , 
\eea
the equations can be written 
\bea 
\Sigma_{n,m} &=& \frac12\left(
\frac{b_{n+1,m+1} - b_{n,m+1} - b_{n+1,m} + b_{n,m}}{k}
+ b_{n+1,m+1}b_{n+1,m} - b_{n,m+1}b_{n,m}
\right)\ , \la{mdkdv1}\\
0 &=& 
\left(\frac{h\Delta _{n, \,m}}{2}  + \frac1{k}  \right)
\left( {b_{n + 1, \,m + 1}} + {b_{n, \,m + 1}} - {b_{n + 1, \,m}} - {b_{n, \,m}} \right)
\la{mdkdv2}\\
&+& 
\left(\frac{h^2\Delta _{n, \,m} }{4k} - \frac1{h}\right) 
\left( {b_{n + 1, \,m + 1}} - {b_{n, \,m+1}}  +{b_{n + 1, \,m}} - {b_{n, \,m }} \right)
\nonumber \\
&+&
\frac{h^2}{4}
\left(
 {\Delta _{n, \,m}}  ({b_{n + 1, \,m + 1}} {b_{n, \,m}} - {b_{n + 1, \,m}} {b_{n, \,m + 1}}) 
- \Delta _{n, \,m}^{2} 
- \Sigma _{n, \,m}^{2} 
\right)
-\frac{h\Delta _{n, \,m}}{k} 
\nonumber\\
&+& 
\frac{h}{4k} 
\left(
b_{n, \,m + 1}^{2}  + b_{n + 1, \,m}^{2} - b_{n, \,m}^{2} - b_{n + 1, \,m + 1}^{2} 
+ 4b_{n + 1, \,m + 1} b_{n, \,m} - 4b_{n + 1, \,m} b_{n, \,m + 1}
\right)
\nonumber\\
&+& 
\frac{h^2}{8k} 
\pmatrix{
b_{n + 1, \,m + 1}^{2} b_{n + 1, \,m} 
- b_{n, \,m}^{2} b_{n, \,m + 1} 
+ b_{n, \,m+ 1}^{2} b_{n, \,m}
- b_{n + 1, \,m + 1}^{2} b_{n, \,m} \cr
- b_{n + 1, \,m}^{2} b_{n + 1, \,m + 1} 
+ b_{n + 1, \,m}^{2} b_{n, \,m + 1}
- b_{n, \,m + 1}^{2} b_{n + 1, \,m} 
+ b_{n + 1, \,m + 1} b_{n, \,m}^{2} \cr
}
\nonumber\\
&+& 
\frac12
\pmatrix{
  3{b_{n + 1, \,m}}\,{b_{n + 1, \,m + 1}} 
+ 3{b_{n, \,m}}\,{b_{n, \,m + 1}} 
-  {b_{n + 1, \,m + 1}}\,{b_{n, \,m}}  \cr
- 2{b_{n, \,m}}\,{b_{n + 1, \,m}} 
-  {b_{n + 1, \,m}}\,{b_{n, \,m + 1}} 
- 2{b_{n + 1, \,m + 1}}\,{b_{n, \,m + 1}} \cr
} \nonumber\\
&+&
\frac{h}{4}
\pmatrix{
  3{b_{n + 1, \,m + 1}} {b_{n + 1, \,m}} {b_{n, \,m + 1}} 
+ 3{b_{n + 1, \,m}} {b_{n + 1, \,m + 1}} {b_{n, \,m}}\cr
- 3{b_{n, \,m + 1}} {b_{n, \,m}} {b_{n + 1, \,m + 1}} 
- 3{b_{n + 1, \,m}} {b_{n, \,m}} {b_{n, \,m + 1}}  \cr
- b_{n + 1, \,m + 1}^{2}  b_{n + 1, \,m} 
- b_{n + 1, \,m}^{2} b_{n + 1, \,m + 1} 
+ b_{n, \,m}^{2} b_{n, \,m + 1}
+ b_{n, \,m + 1}^{2} b_{n, \,m} \cr
} \nonumber\\
&+&
\frac{h^2}{8}
\pmatrix{
b_{n, \,m + 1}^{2} b_{n, \,m}^{2} 
+ b_{n + 1, \,m}^{2} b_{n + 1, \,m + 1}^{2}  
+ 2b_{n + 1, \,m} b_{n + 1, \,m + 1} b_{n, \,m} b_{n, \,m + 1} \cr
- b_{n + 1, \,m}^{2} b_{n + 1, \,m + 1} b_{n, \,m + 1} 
- b_{n + 1, \,m + 1} b_{n, \,m}^{2} b_{n, \,m + 1} \cr
- b_{n, \,m + 1}^{2} b_{n, \,m} b_{n + 1, \,m} 
- b_{n + 1, \,m} b_{n + 1, \,m + 1}^{2} b_{n, \,m}  \cr
} \ .
\nonumber
\eea
Since the field $a$ only appears in the equations through the 
combinations $\Delta$ and $\Sigma$, which only depend on 
$a$ through differences, solutions of this system
are only defined up to  addition of a constant to $a$. 
The analog of propositions 2.1, 3.1 and 4.1 is

\smallskip

\noindent{\bf Proposition 5.1:} 
Let $U_{0,0}(\l)$ be an element of $LGL_2({\bf C})$, 
let $S_{n,m}^{-1}(\l)Y_{n,m}(\l)$ be the Birkhoff decomposition of 
\be
U_{n,m}(\l) = 
\pmatrix{ 1+\frac{h^2\l}{2} & h \cr
            h \l & 1+\frac{h^2\l}{2} \cr}^n
\pmatrix{ 1 & k \cr
            k \l & 1 \cr}^m
   U_{0,0}(\l)\ ,
\ee
and let $a_{n,m}$ and 
$b_{n,m}$ be, respectively, the 1,1-entry  
and  $(-1)$ times the 1,2-entry in the $1/\l$ component of
$S_{n,m}$. Then $a_{n,m},b_{n,m}$ is a solution, possibly with  singularities,  of
the system  \r{mdkdv1}-\r{mdkdv2}. 

\smallskip

The system \r{mdkdv1}-\r{mdkdv2}, despite its algebraic complexity,
is an integrable discretization of 
the equation $b_t=b_x$ in every sense that lattice KdV is. The soliton 
solutions are given as follows:

\smallskip

\noindent{\bf Proposition 5.2:} 
The system \r{mdkdv1}-\r{mdkdv2} has soliton solutions 
\bea 
b_{n,m} &=& \sqrt{\t}\tanh\left(
n \tanh^{-1}\left(\frac{h\sqrt{\t}}{1+\frac12 h^2\t}\right) + m \tanh^{-1}(k\sqrt{\t}) 
+ C
\right)\ , \\
a_{n,m} &=& {\rm constant} ,
\eea
with speed
\be c = \frac{ h \tanh^{-1}(k\sqrt{\t})}
             { k 
   \tanh^{-1}\left(\frac{h\sqrt{\t}}{1+\frac12 h^2\t}\right) }
  \ .\la{newspeed}
\ee

\smallskip

\noindent In greater generality, it can be shown that if instead of equation 
\r{so1}  a $p$th order approximation 
\be
U_{n+1,m} = 
\left[
\sum_{i=0}^p \frac{h^i}{i!}\pmatrix{ 0 & 1 \cr \l & 0 \cr}^i
\right] 
U_{n,m}\ 
\ee
is used, then the speed of the soliton solution is 
\be c = \frac{ h \tanh^{-1}(k\sqrt{\t})}
             { k 
   \tanh^{-1}\left(\frac{s_p(h\sqrt{\t})}{c_p(h\sqrt{\t})}\right) }
  \ ,
\ee
where $c_p(x)$ and $s_p(x)$ are, respectively, the order $p$ truncations 
of the Taylor series for $\cosh(x)$ and $\sinh(x)$ (ignoring terms of order
$x^{p+1}$ and higher). It is straightforward to verify that for small $x$
\be
   \frac1{x} \tanh^{-1}\left(\frac{s_p(x)}{c_p(x)}\right) = 
    \left\{ \matrix{ 1 + O(x^p) & x {\rm ~even} \cr
                     1 + O(x^{p+1}) & x {\rm ~odd} \cr } \right.
\la{tanap}\ee
Thus for small $h$ the dependence of the soliton speed on $h$ becomes weaker
as $p$ increases. Likewise the order of accuracy in 
$k$ can be increased. (The distinction between the even and  odd cases in \r{tanap},
that for odd $p$ there is a ``free'' extra order of magnitude accuracy, means
that \r{mdkdv1}-\r{mdkdv2}, for which $p=2$, 
is actually no more accurate in this regard
than standard lattice KdV, with $p=1$.
The equations obtained from $p=3$ can be written down, but due to their length I 
have restricted the discussion to the $p=2$ case.) 

Returning to the formula \r{newspeed}, note that if $v=k\sqrt{\t}$,
$w=h\sqrt{\t}$, then
\be
c=\frac{\frac1{v} \tanh^{-1}v }{\frac1{w}\tanh^{-1}\left(\frac{w}{1+\frac12 w^2} \right)}\ .
\ee
The function in the numerator increases monotonically from $1$ to $\infty$ as
$v$ goes from $0$ to $1$. 
The function in the denominator decreases monotonically from $1$ to $0$ as
$v$ goes from $0$ to $\infty$. Thus  for the current discretization 
$c$ can only take values greater than 1. 

The soliton solutions just presented can be found using the B\"acklund 
transformation, which is obtained as in previous sections:

\smallskip

\noindent{\bf Proposition 5.3:} If $b_{n,m},a_{n,m}$ is a solution of  
\r{mdkdv1}-\r{mdkdv2} and $\psi_{n,m}$ satisfies 
\bea
&&\frac{\psi_{n+2,m}-2\psi_{n+1,m}+\psi_{n,m}}{h^2}
= 
\t \psi_{n+1,m}\left(1+\frac{h}{4}(b_{n,m}-b_{n+2,m})\right) 
\\
&&~~+\ 
\frac
{ (b_{n+1,m}-b_{n,m})\psi_{n+2,m}
-3(b_{n+2,m}-b_{n,m})\psi_{n+1,m}
+ (b_{n+2,m}-b_{n+1,m})\psi_{n,m}
}{2h}
\nonumber\\
&&~~-\ 
\frac{h^2\t^2}{4}\left(1+\frac{h(b_{n+1,m}-b_{n+2,m})}{2} \right)\psi_{n,m}
+  \frac{a_{n+2,m}-2a_{n+1,m}+a_{n,m}}{2} \psi_{n+1,m}
\nonumber\\
&&~~+\ 
 \frac{(b_{n+2,m}+b_{n,m})b_{n+1,m}-2b_{n,m}b_{n+2,m}}{2} \psi_{n+1,m}
\nonumber\\
&&~~+\ 
 \frac{h}{4} \pmatrix{a_{n,m}b_{n+1,m}  + a_{n+1,m}b_{n+2,m}  + a_{n+2,m}b_{n,m}~~~~~  \cr
          ~~~~~- a_{n,m}b_{n+2,m} -  a_{n+1,m}b_{n,m}  - a_{n+2,m}b_{n+1,m}   \cr} \psi_{n+1,m}
\nonumber
\eea
\bea
\frac{\psi_{n,m+1}-\psi_{n,m}}{k}
&=& 
\frac{ \pmatrix{ \frac{\psi_{n+1,m}-\psi_{n,m}}{h}+ (b_{n+1,m}-b_{n,m+1})\psi_{n,m}~~~~~~~~~~~~~~~~
               ~~~~~~~~~~~~~~~\cr
          ~~~~~~ -\frac{h}{2}\left( \t+a_{n+1,m}-a_{n,m}+b_{n,m+1}(b_{n,m}-b_{n+1,m}) \right)\psi_{n,m} } }
{1+\frac12h(b_{n,m}-b_{n+1,m})}
\eea
then 
\bea
b_{n,m}^{\rm new} &=&  
\frac
{b_{n+1,m}+\frac{\psi_{n+1,m}-\psi_{n,m}}{h\psi_{n,m}} + 
\frac12 h^2(  a_{n,m}-a_{n+1,m}-\t  )}
{1+\frac12h(b_{n,m}-b_{n+1,m})}
\\
a_{n,m}^{\rm new} &=& b_{n,m}b_{n,m}^{\rm new}-a_{n,m}+{\rm constant}
\eea
is also a solution of \r{mdkdv1}-\r{mdkdv2}. 

\smallskip

\noindent All formulae in the previous proposition have been written in a 
manner that hopefully makes it clear in what sense they are modifications 
of the corresponding formulae in proposition 3.2. The surprising feature 
of the discretization of the Schr\"odinger equation in proposition 3.2, 
equation \r{dsl1}, is that in it the parameter $\t$ multiplies $\psi_{n,m}$,
not $\psi_{n+1,m}$, which would seem more natural. The new discretization 
just presented, equation \r{mdkdv1}, has $\t$ multiplying $\psi_{n+1,m}$.
But the cost of this is the introduction of many new terms, including 
a term proportional to $\t^2$, multiplying $\psi_{n,m}$. It can be checked 
that the new discretization \r{mdkdv1} is a second order approximation to 
the Schr\"odinger equation, while \r{dsl1} is only first order. This is 
the justification for the title of this section.

\section{Concluding Remarks}

In this paper I have presented a systematic approach towards integrable
discretizations, based on the loop group approach to integrable systems.
Three integrable discretizations have been examined in detail, one known,
the lattice KdV system of Nijhoff {\em et al.}, and two new, one of 
I have called {\em full lattice KdV}, as it would seem to be 
the first discrete integrable system with (potential) KdV as its standard 
continuum limit. For each integrable discretization 
a B\"acklund transformation has been given and 
soliton solutions have been analyzed. Unlike the 
lattice KdV system of Nijhoff {\em et al.}, which only displays solitons 
with speeds below, above or equal to 1, full lattice KdV has the full range
of soliton speeds (for suitable choices of $h$ and $k$).

Full lattice KdV would seem to merit further attention. Our plans 
include conducting numerical studies, and to try to work out a suitable 
inverse scattering formalism. Another issue that has not been touched 
upon in this paper is the subject of tau functions for discretizations. 
The linear flows on a loop group that underlie KdV can be extended
to the central extension of the group, and one would expect the same 
to be true for the discretizations looked at in this paper. 

The formalism developed here can also be extended to look at 
integrable discretizations of KdV on non-rectangular lattices, see \c{hex}.

\section*{Acknowledgments}

I would like to thank David Kessler for discussions. This work was supported
by the Israel National Science Foundation.

\end{document}